Jacek DAMEK[*]

# THE SCALING LIMITS OF KMS STATES ON THE RINDLER HORIZON

*Abstract*

*The standard concept of scaling limits of distributions on manifolds is reformulated, and then a new framework for scaling at boundary points is provided. Next, we introduce a class of so-called $L^1$ - KMS states, which is subsequently fully characterized for the Rindler wedge. Using these tools, we compute rigorously the scaling limit of the regular β-KMS state of the free scalar quantum field on the Rindler horizon. Thereby, we correct certain inaccurate results of founding paper [1], nevertheless fully corroborating the physical essence of them.*

## 1. INTRODUCTION

The first part in the process of quantization of a linear field seems to be, in general, indisputable: one takes the Weyl algebra over classical solutions of a wave equation as (at least) model observable algebra, and imposes the time evolution corresponding to the time-like Killing vector on the space-time. Second, one has to choose a set of physical states on the algebra in order to get numerical predictions, as well as to build Hilbert space structures, usually via the famous GNS construction. At this very point, a considerable problem has emerged since the first papers on the subject [see, e.g., 13, and 14]. One is unable to pick out a preferred folium of states on the grounds of evident and unquestionable ideas. Even the existence of such a unique folium is doubtful [3]. Since a folium (in a stationary case) is expected to be picked out as, for instance, the normal states with respect to some time-invariant quasi-free 'vacuum' state [14], we will discuss that possibility.

In the past two decades the vast literature [e.g., 4, 15] has been devoted to the so-called Hadamard states, defined by the singularity structure of their two-point function being (informally)

$$G_{sing}(x, y) = 1/(2\pi)^2 \left( \Delta^{1/2}/\sigma + v \ln \sigma \right), \qquad (1.1)$$

---

[*] Institute of Mathematics, Technical University of Zielona Góra, Poland
The author supported by the Individual Grant of the Rector of Technical University of Zielona Góra.

where $\sigma(x, y)$ denotes the squared geodesic distance between $x$ and $y$. $\Delta$, $v$ are some $C^\infty$ functions on the space-time considered (see, [4], for a rigorous definition).

However, the number of constructive outcomes on Hadamard states appears to be limited. One of the general existence results is that of Fulling, Narcowich and Wald [15] for a class of ground states on static space-times.

Another regularity condition to be imposed on states has been proposed and discussed in [1, 2], and we shall later refer to it as the Scaling Limit Condition (SLC). This stipulation is substantially weaker than the Hadamard one, but occurs in applications to yield the same vacuum states and its temperatures. Moreover, it can often be tractable constructively. We get SLC from (1.1) taking its scaling limit (defined in [2] and here). Thus we obtain the required form of this limit, denoted by slim, for physical vacua:

$$\text{slim} \, w^{(2)} = (2\pi)^{-2} (g_{\mu\nu} z^\mu z^\nu)^{-1} \qquad (1.2)$$

The most stringent and fruitful demanding related to (1.2) is to postulate it on a horizon of a space-time, as we do in Sec. 4.

Having regard to all the above-noted, we present this article as a forerunner of the series of papers [3, 9] aimed at the study of interconnections between the Hadamard condition and SLC in curved space-times and their physical meaning. In [3] we evaluate the scaling limits of the two-point functions of the regular KMS states and check SLC for a simple class of space-times by straightforward means used in [1]. Unfortunately, there are certain mistakes in that paper, but we mend them developing rigorous theory. Paper [9] relies on operator theory and gives explicit forms of scaling limits of quantum fields, thus SLC, on stationary space-times with Killing horizons [cf. 4].

In this contribution, we only report certain results of [3] leading to constructive theorems on SLC in the Rindler wedge.

## 2. SCALING LIMITS OF DISTRIBUTIONS

We begin with a brief review of well-known basic definitions and properties of scaling limits [2], but adapted to our physical aims. Assume $\Omega$ to be an open, and (without loss of generality) including null vector, subset of $\mathbf{R}^n$.

**2.1. Definitions.** Let $\mathbf{R}^+$ be the positive numbers.

(a) The *scaling function* will be any function $N: (0, a) \to \mathbf{R}^+$, where $a>0$ may depend on $N$. $E$ is the class of those functions.

(b) For a complex-valued function $f$ and $\lambda>0$ we write
$$f_\lambda(x) = f(\tfrac{x}{\lambda}) \quad (x \in \mathbf{R}^n).$$

(c) If $u \in D'(\Omega)$, $N \in E$ we may define a mapping $\text{slim}(u; N): D(\mathbf{R}^n) \to \mathbf{C}$ by $\text{slim}(u; N)(\varphi) = \lim_{\lambda \to 0} N(\lambda) \langle u, \varphi_\lambda \rangle$, whenever the limits exist, and call it the *scaling limit of a distribution u (w.r.t. N)*.

**2.2 Proposition.** slim(*u; N*) *is a distribution on* **R**$^n$.

Recall that a distribution $u \in D'(\Omega)$, is called homogenous if $u_\lambda = \lambda^\alpha u$, where $\langle u_\lambda, \varphi \rangle = \langle u, \varphi_\lambda \rangle$, and $\alpha \in$ R.

**2.3. Proposition.** slim(*u; N*) *is a homogenous distribution (thus tempered).*

For convenience of the usage of an unequivocal notion of a scaling limit, independent from a scaling function to be chosen, we restrict ourselves to so-called *scaled distributions,* defined to be those possessing some non-zero scaling limit. The following proposition asserts that for a scaled distribution its scaling limits differ only by scale factors.

**2.4. Proposition.** *Let* $N, S \in E, u \in D'(\Omega)$ *and* $\text{slim}(u; N) = u_0 \neq 0$.

*u possesses a scaling limit v ≠ 0 w.r.t. S if and only if there exists α>0 such that*

$$\lim_{\lambda \to 0} \frac{S(\lambda)}{N(\lambda)} = \alpha.$$

*Then,* $v = \alpha u_0$.

Thus, we come to the well-defined absolute notion of a scaling limit.

**2.5. Definition.** Let $u \in D'(\Omega)$ be a scaled distribution, $N \in E$ and $\text{slim}(u; N) \neq 0$. The scaling limit of *u,* denoted by slim(*u*), will be defined to be [slim(*u*)], where [*v*] denotes the set $\{\alpha v : \alpha > 0\}$.

Obviously, it is a partition into classes of equivalence.

Now, we will generalize the notion of a scaling limit to the case of scaling at a boundary point of the region, where a distribution is defined. This concept enables one to deal rigorously with the scaling of distributions of physical interest on space-time horizons, particularly on the Rindler horizon.

In what follows we assume the zero point to belong to the closure of $\Omega$. The omitted for brevity proofs, throughout the rest of the paper, can be found in [3], unless otherwise stated.

**2.6. Definition.**

Let *V* be a region in **R**$^n$. If for each compact $K \subset V$ there exists a number $\lambda_0 > 0$ such that the inequality $0 < \lambda < \lambda_0$ implies the inclusion $\lambda K \subset \Omega$, then we say that the set *V* is *contractible (to zero) through the region* $\Omega$. The class of all such sets *V* we denote by $\mathfrak{C}(\Omega)$.

It is clear that if $u \in D'(\Omega)$ then the *V* is an admissible domain of potential exis-tence of a scaling limit of *u,* as scaling $\lambda \to f_\lambda$ does pull the support of *f* into Ω. However, the question arises when such a domain can exist. It is answered by the subsequent simple criterion involving the notion of conical regularity, which we shall now expose.

**2.7. Definitions.** (a) A cone in $\mathbf{R}^n$ (n>1) is a set that is isometric to the set

$$S_0 = \left\{ x \in R^n : \sqrt{x_1^2 + \cdots x_{n-1}^2} < \tfrac{r}{h} x_n, x_n < h \right\} \quad \text{for some } r, h>0.$$

For **R**, a cone is a bounded open interval.

(b) Given $A \subset R^n$, we write ex*A* to denote the set $\left\{ \lambda x : x \in A, \lambda \in R^+ \right\}$.

**2.8. Definition.** (the internal cone property)

Let $p \in \overline{\Omega}$. If there exists a cone with the apex at *p* and included in Ω, then we call *p* a *conically regular point of* $\overline{\Omega}$.

The next theorem provides convenient criteria to assure which points of $\overline{\Omega}$ fulfil the above condition.

**2.9. Theorem.**

(1) *If Ω is convex, then every boundary point of this set is conically regular.*

(2) *Let $\Omega \subset R^n, n > 1$. If the boundary of Ω is piecewise smooth, then the regular points of this boundary are conically regular.*

Finally, we are prepared to state the above-promised criterion for contractibility. All cones considered further will have their apexes at 0.

**2.10. Theorem.** *The existence of a contractible (through Ω) region is equivalent to the conical regularity of the zero point.*

The following facts will recover the nature of contractible sets and show the maximal region in which scaling can be expected to make sense.

**2.11. Theorem.** *Suppose zero to be a conically regular point of $\overline{\Omega}$. Then the family $\mathfrak{C}(\Omega)$, ordered by inclusion, has the maximal element $\mathfrak{M}(\Omega)$, which takes the form*

$$\mathfrak{M}(\Omega) = \bigcup_{S \in C} \mathrm{ex} S,$$

where *C* stands for the cones included in Ω if 0 belongs to Fr Ω, otherwise $\mathfrak{M}(\Omega) = \mathbf{R}^n$.

$\mathfrak{M}(\Omega)$ will be referred to as the *maximal contractible* (through *Ω*) *region.*

Now, we are able to present a desired definition of a scaling limit of a distribution in the general case of $0 \in \overline{\Omega}$.

**2.12. Definition.** Let *V* be a contractible through *Ω* region in $\mathrm{R}^n$, $u \in D'(\Omega), N \in E$.

Assume that the limits $\lim_{\lambda \to 0} N(\lambda) \langle u, \varphi_\lambda \rangle$ exist for all $\varphi \in D(V)$.

Then one can define slim($u; N$) as a distribution on $V$ as previously:

$$\text{slim}(u; N)(\varphi) = \lim_{\lambda \to 0} N(\lambda) \langle u, \varphi_\lambda \rangle \quad (\varphi \in D(V)).$$

It is a matter of simple work to check that all definitions and properties in this section hold for such a bit more general scaling limits after obvious modifications. In the end of the present section, we shall outline the natural extension of the hitherto obtained results to general manifolds. Since a manifold is locally identical to its tangent space, we have only to rewrite the results of the previous considerations in terms of distributions on a manifold to obtain a desired theory. The problem remains only in proving the independence of the construction and properties from a choice of an identification mapping.

From now on, we assume that $M$ is a $C^\infty$ paracompact manifold, $\Omega$ - an open subset of $M$, and $p \in \overline{\Omega}$.

**2.13. Definition.** Let $q \in M$, and $\Phi$ be a $C^\infty$ - diffeomorphism, which maps some neighborhood (open) of 0 in $T_q M$ onto the one of $q$, so that

$$\Phi(0) = q, \quad \Phi'(0) = \text{id} \quad (T_0 T_q M \approx T_q M).$$

Then $\Phi$ will be called an *identification mapping between $T_q M$ and $M$*.

**Example.** Let $\psi: V \to W = \psi(V)$ be a coordinate chart, $\psi(q) = x$.

Then $\Phi(\alpha_i \partial_i) = \psi^{-1}(x + \alpha)$ is a standard identification mapping.

The definitions concerning with the regions of the existence of a slim($u$) (contractibility !) are reduced to those on $\mathbf{R}^n$ by passing to $\mathbf{R}^n$ from $\Omega \subset M$ on employing the mapping $(\Phi \circ I)^{-1}$, where I is a linear isomorphism between $T_p M$ and $\mathbf{R}^n$.

These notions are not dependent on a specific choice of $\Phi$ or I, because neither $\Phi_2^{-1} \circ \Phi_1$ nor $I_2^{-1} \circ I_1$ can change the feature of the existence of a suitable cone, and for $(\Phi_2^{-1} \circ \Phi_1)' = \text{id}$ a change of $\mathfrak{M}(\Omega)$ viewed as included in $T_p M$ is not possible, either.

After slight modifications theorems 2.9 - 2.11 are still valid.

Now we shall focus on a definition of a scaling limit and a statement of its $\Phi$ - independence adopting the preceding notation

**2.14. Definition.** We define $\Phi_* : D(V) \to D(W)$ by

$$\Phi_* f(q) = f(\Phi^{-1}(q)), \quad \text{for each } f \in D(V).$$

Such $\Phi_*$ is a topological linear isomorphism.

Let $U \subset T_p M$ be a contractible through $\Omega$ region, and let $u \in D'(\Omega), N \in E$. As-sume that the limits $\lim_{\lambda \to 0} N(\lambda) \langle u, \Phi_* \varphi_\lambda \rangle$ exist for all $\varphi \in D(U)$.

**2.15. Definition.** One can define $\text{slim}_p(u; N; \Phi)$ to be a distribution on $U$:

$$\langle \text{slim}_p(u; N; \Phi), \varphi \rangle = \lim_{\lambda \to 0} N(\lambda) \langle u, \Phi_* \varphi_\lambda \rangle \quad (\varphi \in D(U)).$$

**2.16. Theorem** [2, 3]. *Under the above assumptions, but with two identification mappings* $\Phi_1$, $\Phi_2$, *it follows that, if* $\text{slim}_p(u; N; \Phi_1)$ *exists on U, then so does* $\text{slim}_p(u; N; \Phi_2)$, *and both the limits are equal*.

Consequently, we may dispose of the apparent dependence on $\Phi$, and simply write $\text{slim}_p(u; N)$, or $\text{slim}_p(u)$ for a class, as before.

In Section 4 we shall need the following coincidence limit, defined by $\text{slim}^c_p(u) = \text{slim}_{(p,p)}(u)$, where $u \in D'(\Omega \times \Omega)$ and $p \in \overline{\Omega}$. Its domain of existence is characterized by evident

**2.17. Proposition.** $(p, p)$ *is a conically regular point of* $\overline{\Omega \times \Omega}$ *if and only if so is p for* $\overline{\Omega}$. *Then it follows that*

$$\mathfrak{M}_{(p,p)}(\Omega \times \Omega) = \mathfrak{M}_p(\Omega) \times \mathfrak{M}_p(\Omega).$$

Summing up, all the results of this section also hold mutatis mutandis for manifolds.

## 3. $L^1$ - STRONGLY CLUSTERING KMS STATES

Our explicit derivation of the scaling limit of KMS states in the Rindler wedge is feasible due to formula (Prop. 3.6), informally stated in [1]. Therefore, we will presently introduce a class of KMS states (Def. 3.1) for which this formula can only hold, and provide a variety of theorems on the actual contents of the class, so as to find the range of our calculations in Sec. 4. We are exclusively interested in the linear theory in a curved space-time, thus we have stopped on this level of generality, though the majority of the results may be formulated for generic Weyl algebras (and even not Weyl ones).

Let $(M, g)$ be a $C^\infty$ paracompact, stationary, globally hyperbolic space-time of the dimension $n$ with the orientable Cauchy surface. Consider a classical evolution equation:

$$(\nabla^a \nabla_a + V)\varphi = 0, \quad \text{where } V \in C^\infty(V) \tag{3.1}$$

One may construct [4, 5] the Weyl algebra $A$ on the simplectic space $(D, \sigma)$ of real $C^\infty$ solutions of (3.1) (having compact support on Cauchy surfaces), where $\sigma$ is a standard bilinear symplectic form for eq. (3.1). We shall be passing from our '$\varphi$-expressions' to $D(M)$ via the advanced-minus-retarded fundamental solution of (3.1) denoted by $E$, e.g., $\varphi(f) \equiv \varphi(Ef)$, where $\varphi$ stands for a quantum field [4, 5]. This amounts to the usual $n$-dimensional smearing of a pointwise-defined field, $\varphi(x)$. The classical symplectic evolution $\varphi \to \varphi_t$ gives rise to the group of automorphisms $\{\tau(t): t \in R\}$ on the algebra $A$:

$$\tau(t)[W(\varphi)] = W(\varphi_t).$$

Therefore, $(A, \tau)$ forms a $C^*$-dynamical system, which will be a basis for our further considerations. For simplicity, we adopt the convention that a ground state is a $\beta$-KMS state with $\beta = \infty$, and regard only states with a zero one-point function, as covering the non-zero case is simple, but troublesome [3].

**3.1. Definition.** A $C^2$ state $\omega$ over $(A, \tau)$ is defined to be $L^1$-*strongly clustering* (or $L^1$-*state*) if the function $t \to \omega(\varphi(f)\tau_t[\varphi(g)])$ is integrable for any $f, g \in D(M)$.

Rich interconnections among assumptions of that type [see, e.g.,6] are discussed in [3]. The next proposition is rather auxiliary.

**3.2. Proposition.** *Let $\omega$ be a quasi-free $\tau$-invariant $L^1$-state over $(A, \tau)$. Then it is regular ("no zero-modes").*

*Proof.* Consider the one-particle Hilbert structure of $\omega$, namely $(K, H, U(t))$ [4, p.78]. For an arbitrary unit $x \in H$ one may find $y \in H$ such that

$$\|x - y\| < \tfrac{1}{4}, \quad y = K\varphi + iK\varphi' \text{ for some } \varphi, \varphi' \in D \qquad (3.2)$$

as $KD + iKD$ is dense in $H$. Then

$$\langle x|Ux \rangle = \langle y|Uy \rangle + \langle y|U(x-y) \rangle + \langle x-y|Ux \rangle \quad \text{and} \quad \|y\| \le \|x-y\| + \|x\| = \tfrac{5}{4}. \text{ Hence,}$$

$$|\langle x|U(t)x \rangle| \le |\langle y|U(t)y \rangle| + \tfrac{9}{16}. \qquad (3.3)$$

Note that $t \to |\langle y|U(t)y \rangle|$ is an integrable function by (3.2) and $L^1$-strong clustering of $\omega$. It then follows from (3.3) that $|\langle x|U(t_0)x \rangle| < 1$ for some $t_0$, and thus $x$ is not stable under the group $U$, though it is arbitrary. ∎

**3.3. Theorem.** *Let $\omega \in C^2$ be an $(L^1, \beta)$-KMS state over $(A, \tau)$. It follows that $\omega$ is a quasi-free state.*

*Proof.* We can only give here a reasoning in the case of a ground state, deferring the more complicated $\beta < \infty$ case to [3].

Given the state $\omega_\infty$, we may construct its liberation, i.e., the quasi-free state $\omega^L_\infty$ with the same two-point function. If

$$\omega_\infty(\varphi(\varphi_1)\varphi(\varphi_2)) = \mu(\varphi_1, \varphi_2) + \tfrac{1}{2}i\sigma(\varphi_1, \varphi_2), \qquad (3.4)$$

then

$$\omega_\infty^L(W(\varphi)) = \exp\left(-\tfrac{1}{2}\mu(\varphi,\varphi)\right). \tag{3.5}$$

Let (K, H) be the one-particle structure for $\omega_\infty^L$. By (3.4) and (3.5) the state $\omega_\infty^L$ is $\tau$-invariant, so the unitary $U(t)$ performs the time evolution in H. Moreover, the group $t \to U(t)$ is strongly continuous as

$$\langle K\varphi_1 | U(t) K\varphi_2 \rangle = \omega_\infty^L(\varphi(\varphi_1)\varphi(\varphi_{2t})) = \omega_\infty(\varphi(\varphi_1)\varphi(\varphi_{2t})), \tag{3.6}$$

the last term is continuous by assumption, and the complexified range of K is dense in H. Therefore, $U(t) = \exp(ith)$ for some self-adjoint h. If one had $h \geq 0$, the state $\omega_\infty^L$ would also be ground, but that stems from the analyticity properties implied by (3.6). Namely, $t \to \langle K\varphi_1 | \exp(ith) K\varphi_2 \rangle$ has the analytic extension to the upper complex plain, which is bounded, since, by (3.6), it is equal to the two-point function of $\omega_\infty$ that certainly possesses these characteristics on representing it in the GNS construction. This suffices to proving that $h \geq 0$ [6, 3]. In addition, $\omega_\infty^L$ is an $L^1$-state by (3.6), hence, by Proposition 3.2, it is regular. However regular and ground quasi-free states must be pure [4]. Finally, invoking the theorem due to Kay [7], asserting that a state on $A$ is equal to its liberation if this latter state is pure, finishes the proof. ∎

**3.4. Corollary.** *The $L^1$-KMS states are included in the regular quasi-free states. Thereby, for each $\beta \leq \infty$ there exists at most one such a state.*

The next theorem illustrates how large a family of $L^1$-KMS states may be.

**3.5. Theorem.** *Let $\omega_{\beta_0}$, $\beta_0 < \infty$ be an $(L^1, \beta_0)$-KMS state over $(A, \tau)$. If either of the following conditions is fulfilled:*

*(a) for any $\varphi, \varphi' \in D$ there exists an integrable function M such that $g \in L^1$, $|g'(t)| \leq M(t)$, and $M(t) \to 0$ monotonously as $t \to \infty$ or $t \to -\infty$,*

*where $g(t) = \omega_{\beta_0}(\varphi(\varphi)\varphi(\varphi'_t))$ $(g$ is always $C^\infty$, [3]);*

*(b) for any $\varphi, \varphi' \in D$ there exist $C>0$, $\alpha > \sqrt{3}$ such that*

$$|g(t)| \leq \frac{C}{1+|t|^\alpha},$$

*then for each $\beta \leq \infty$ there exists a (unique) $(L^1, \beta)$-KMS state over $(A, \tau)$.*

*Proof.* A rather lengthy proof is contained in [3].

The proposition below is crucial for our method of computations in Sec. 4, as it expresses the two-point function of an $L^1$-KMS state in terms of the well-known commutator function.

**3.6. Proposition.** *Let $\langle \,\cdot\, \rangle_\beta$ be an $(L^1, \beta)$-KMS state over $(A, \tau)$. Then, for all $f, g \in D(M)$,*

$$\langle \varphi(f)\varphi(g) \rangle_\beta = \frac{1}{2\pi} vp \int_{-\infty}^{\infty} \frac{1}{1-e^{-\beta\omega}} \int_{-\infty}^{\infty} \langle [\tau_t(\varphi(f)), \varphi(g)] \rangle_\beta \, e^{i\omega t} dt\, d\omega. \quad (3.7)$$

*Proof.* First, note that in the KMS condition written as

$$\int_{-\infty}^{\infty} dt\, f(t) \langle AB_t \rangle_\beta = \int_{-\infty}^{\infty} dt\, f(t+i\beta) \langle B_t A \rangle_\beta, \quad (3.8)$$

with $A = \varphi(f)$, $B = \varphi(g)$, $\hat{f} \in D$, valid for quasi-free states, one can set $f(t) = \exp(it\omega)$ by the Lebesgue-dominated convergence theorem and $L^1$ property of $\langle \,\cdot\, \rangle_\beta$. Next, one readily derives the desired formula from (3.8) with this setting. ∎

*Remark.* The limit $\beta \to \infty$ in (3.7) gives the relationship for ground states.

## 4. THE SCALING LIMITS OF $L^1$-KMS STATES ON THE RINDLER HORIZON

Now, we will specify to the model case of the Rindler wedge. Thus, our space-time is $W_r = \{x \in Min : x^1 > |x^0|\}$, where *Min* stands for usual four-dimensional Minkowski space. The time evolution $\tau$ is set by Lorentz boosts. We consider fields interacting only with gravity, i.e., $V(x) = m^2$, $m \geq 0$ in wave equation (3.1).

**4.1. Proposition.** *The family of $L^1$-KMS states over $(A(W_r), \tau)$ consists of all regular quasi-free KMS states. Actually, it is the Fulling-Kay family of states constructed rigorously in* [8] ([cf. 16]).

*Proof.* In view of Theorem 3.3 we need only to show that the Fulling-Kay states are $L^1$. Regard the $2\pi$-KMS state of the former type, $\omega_{2\pi}$. From the Bisognano-Wichmann theorem, it is a restriction to $W_r$ of the usual ground state $\omega_0$ on *Min*. Thus, $\omega_{2\pi}(\varphi(f)\tau_t[\varphi(g)])$ is the two-point function of $\omega_0$ expressed in the Rindler time coordinate. As the explicit form of this function is known, one readily sees that it is a L. Schwartz function. On applying Theorem 3.5, Corollary 3.4, we may claim the desired statement. ∎

*Remark.* It turns out that $\bigvee_{\beta < \infty} \omega_\beta \in S$. ($S$ = the states with a L. Schwartz two-point function, cf. Def. 3.1)

Before exhibiting the explicit shape of the scaling limit of a KMS regular state in $W_r$,

we need some preparatory work.

Let $H^+$ and $H^-$ be the future and past event horizon, respectively. If $p \in H^+ \cap H^-$, then $p$ is clearly conically regular (we identify $T_p Min = Min$) point of $\overline{W}_r$ (Theorem 2.9), $\mathfrak{M}_p(W_r) = W_r$. Therefore, by Prop.2.17 and definition above it, we may treat $\text{slim}_p^c$ of a distribution in $W_r \times W_r$ as a distribution in the same space.

Let $\omega_{\beta,m}^{(2)}$ denote the two-point function of the β-KMS regular state $\omega_{\beta,m}$ on $W_r$ with value $m$ of mass.

**4.2. Theorem.** *Let* $p \in H^+ \cap H^-$. *Then*

$$\text{slim}_p^c \left( \omega_{\beta,m}^{(2)} \right) = \omega_{\beta,0}^{(2)}.$$

*Proof.* It is based on relationship (3.7), where the Pauli-Jordan function substitutes for the commutator in the right-hand side of the formula, and the following lemma.

**4.3. Lemma.** *Let* $\varphi, \hat{\varphi} \in L^1$. *It follows that*

$$vp \int_{-\infty}^{\infty} \frac{1}{1 - e^{-\beta\omega}} \int_{-\infty}^{\infty} \varphi \, e^{i\omega t} dt d\omega = \frac{i\pi}{\beta} \lim_{\varepsilon \to 0} \int_{-\infty}^{\infty} \coth\left( \frac{\pi}{\beta}(t + i\varepsilon) \right) \varphi \, dt$$

**4.4. Corollary.** $\omega_{\beta,m}$ *satisfies the Scaling Limit Condition if and only if* $\beta = 2\pi$.

*Proof.* Only for $\beta = 2\pi$ has the explicitly known $\omega_{\beta,0}^{(2)}$ [17] the form required in SLC (Eq. 1.2).

## 5. DISCUSSION.

The case of the Rindler wedge is rather trivial, however one can enlarge the class of space-times for which Theorem 4.2 holds by conformal transformations applied to $W_r$. As the formal multiplication of Green functions by conformal factors could not correspond to actual KMS states, one may use reconstruction (from Green functions) theorems of [10, 11, 12]. In this way, we get the result, for instance, for the four-dimensional model of the Schwarzschild black hole yielding the proper value of the Hawking temperature [3]. Alternative, more general methods being used in [9] give the analogue of Theorem 4.2 for a rich class of curved space-times. Therefore, occasionally, our results (here in the case of Rindler) have fully corrected certain inaccurate computations in [1], where the final form of $\text{slim}_H(w^{(2)})$ is proportional to $\beta^{-1}$ (nevertheless yielding the true Hawking temperature !). However, that is impossible whatsoever linear definition of slim we adopt. In particular, if $\text{slim}\left(\omega_\beta^{(2)}\right) = \beta^{-1}v$, $\Delta$ - Pauli-Jordan func-tion, then $\langle \text{slim}(\Delta), f \otimes g \rangle = \langle \text{slim}\omega_\beta^{(2)}, f \otimes g \rangle - \langle \text{slim}\omega_\beta^{(2)}, g \otimes f \rangle = \beta^{-1} v' \neq 0$, in contradiction with the fact that the left-hand side of this equality is independent from $\beta$.

A more comprehensive physical discussion on the connections between Hadamard states and SLC can be found in [3].

*Acknowledgements.* The author thanks Professor Roman Gielerak for helpful discussions on the subject.

## REFERENCES

1. R. Haag, H. Narnhofer and U. Stein, Commun. Math. Phys. **94** (1984) 219.
2. K. Fredenhagen, R. Haag, Commun. Math. Phys. **108** (1987) 91
3. J. Damek, *L¹-Strongly Clustering KMS States and their Scaling Limits on Space-times with a Horizon,* in preparation.
4. B.S. Kay, R.M. Wald, Phys. Rept. **207** (1991) 49.
5. J. Dimock, Commun. Math. Phys. **77** (1980) 219.
6. O. Bratelli, D.W. Robinson, *Operator Algebras and Quantum Statistical Mechanics, II,* Springer, Berlin, 1981.
7. B.S. Kay, J. Math. Phys. **34** (1993) 4519.
8. B.S. Kay, Commun. Math. Phys. **100** (1985) 57.
9. J. Damek, *On Existence of KMS Hadamard States in Curved Space-times*, in preparation.
10. R. Gielerak, L. Jakóbczyk, R. Olkiewicz, J. Math. Phys. **39** (1998), 6291.
11. R. Gielerak, L. Jakóbczyk, R. Olkiewicz, J. Math. Phys. **35** (1994) 3726.
12. R. Gielerak, L. Jakóbczyk, R. Olkiewicz, J. Math. Phys. **35** (1994) 6291.
13. S.A. Fulling, Phys. Rev. **D7** (1973) 2850.
14. R. Haag, Local Quantum Physics, Springer-Verlag, 1991.
15. S.A. Fulling, F.J. Narcowich, R.M. Wald, Ann. Phys. **136** (1981) 243.
16. S.A. Fulling, S.N.M. Ruijsenaars, Phys. Rep. **152** (1987) 135.
17. J.S. Dowker, Phys. Rev. **D18** (1978) 1856.